\documentclass[conference]{IEEEtran}
\IEEEoverridecommandlockouts
\usepackage{cite}
\usepackage{amsmath,amssymb,amsfonts}
\usepackage{algorithmic}
\usepackage{graphicx}
\usepackage{textcomp}
\usepackage{xcolor}
\usepackage{multirow}
\usepackage{subcaption}
\begin{document}

\title{Federated Learning for Wireless Applications: A Prototype \thanks{We acknowledge the research grants from MeITY and  DST SERB, the govt. Of India.}
}
\author{\IEEEauthorblockN{Varun Laxman Muttepawar$^{1,*}$, Arjun Mehra$^{1,*}$, Zubair Shaban$^{*}$, Ranjitha Prasad$^{*}$, J. Harshan$^{\dagger}$}
\IEEEauthorblockA{$^{*}$Indraprastha Institute of Information Technology Delhi, India; $^{\dagger}$Indian Institute of Technology Delhi, India} \vspace{-3mm}
}

\maketitle

\begin{abstract}

Wireless embedded edge devices are ubiquitous in our daily lives, enabling them to gather immense data via onboard sensors and mobile applications. This offers an amazing opportunity to train machine learning (ML) models in the realm of wireless devices for decision-making. Training ML models in a wireless setting necessitates transmitting datasets collected at the edge to a cloud parameter server, which is infeasible due to bandwidth constraints, security, and privacy issues. To tackle these challenges, Federated Learning (FL) has emerged as a distributed optimization approach to the decentralization of the model training process. In this work, we present a novel prototype to examine FL's effectiveness over bandwidth-constrained wireless channels. Through a novel design consisting of Zigbee and NI USRP devices, we propose a configuration that allows clients to broadcast synergistically local ML model updates to a central server to obtain a generalized global model. We assess the efficacy of this prototype using metrics such as global model accuracy and time complexity under varying conditions of transmission power, data heterogeneity and local learning.

\end{abstract}
\vspace{-2mm}
\section{Introduction}

In the era of global connectivity and the Internet of Things (IoT), efficient and privacy-preserving machine learning (ML) solutions are crucial. Federated Learning (FL) \cite{b1}, an innovative paradigm in ML, emerges as a powerful approach to train models across decentralized devices without compromising on data privacy. Essentially, FL advocates the aggregation of local ML model parameters to obtain a global ML model. 

In this dynamic landscape of modern connectivity, where wireless devices permeate our daily lives, the significance of implementing FL on hardware, for example, using software-defined radios (SDRs), is apparent. SDRs serve as the backbone of decentralized networks, facilitating real-time communication. Harnessing the power of FL on such hardware not only optimizes model training but also ensures that privacy-preserving ML becomes a tangible reality. 
The key objectives of our work are the following:
\begin{itemize}
    \item Privacy-Preserving ML: Establishing an FL protocol, which is privacy-preserving by design and transmits parameter updates over the wireless channel.
    \item Real-time ML: Demonstrating the seamless integration of learning with NI-USRPs (National Instruments-Universal Software Radio Peripheral) and Zigbee for real-time update communication.
    \item Decentralized Model Training: Showcasing the efficiency gains achieved by distributing the learning process across the hardware devices.
\end{itemize}
Our demo provides a hands-on guide for setting up an FL environment using USRPs and Zigbee devices. The convergence of FL principles with tangible hardware platforms opens new avenues for collaborative and privacy-conscious ML, thereby unlocking the potential of FL on wireless devices. This work also helps to bridge the gap between theoretical advancements \cite{b2,b3} and practical implementations in wireless FL.
\vspace{-2mm}
\section{Federated Learning \& FedAvg Algorithm}

Federated learning is a decentralized ML approach where model training occurs locally on the edge devices, and only the model updates, not the raw data, are shared with a central server. In particular, in Federated Averaging (FedAvg), clients share their parameter updates with the server. The server aggregates these updates by computing a weighted average which is assigned as a global ML model. The algorithm is summarized in the following steps:
\begin{enumerate}
\item Initialization: The central server initializes and broadcasts a global initialization to the clients.
\item Training \& Local updates: Each client computes the local updates by training the model on its local data and transmits them to the central server.
\item Aggregation \& Global model: The central server aggregates the local updates using a weighted average and updates the global model.
\item Re-transmission of Global model: The central server then broadcasts the new global model to all the clients, who train on it for the next round.
\item Convergence criteria are monitored, such as reaching an accuracy threshold or a number of iterations.
\end{enumerate}
This collaborative training paradigm enables privacy-preserving communication, making it particularly suitable for scenarios with sensitive distributed data sources, such as mobile devices or edge computing environments. Each edge device is tasked with locally training a model based on its received signal data, capturing unique characteristics of the wireless environment. The goal is to create a global model that not only performs well on the individual signals observed by each edge device but also generalizes effectively across the diverse signal characteristics contributed by different devices. One such work is studied in \cite{b4}, where the goal is to train models that can identify adversarial attacks in a decentralized fashion by exploring the usage of deep neural networks. The intent is to create a strong and secure wireless network where numerous nodes work together to detect and eliminate attacks by implementing the principles of FL.

\section{Details of the Demonstration}
In this section, we discuss the details of the proposed demonstration. Our experimental setup is based on the federated averaging protocol as described in the previous section. Throughout the training process, only model updates (in the form of gradients) are exchanged with a central server, thereby preserving local raw data. Through iterative updates and aggregation, a global model emerges,  embodying generalized features of the data despite each client having observed only a subset of the data. We conducted the experiments on a reduced MNIST dataset, where we consider any $4$ classes distributed across two clients. We simulate the following scenarios:
\begin{itemize}
    \item Independent and identically distributed (IID) Setup: We shuffle the dataset and randomly assign half of the dataset to each client. Since we assign the samples randomly, the local data at the clients are inherently IID. 
    \item Non-IID setup: We assign two classes to each of the clients with no overlap, i.e., the client's local dataset is sampled from different distributions and, hence, non-IID. 
\end{itemize}

\subsection{Local Learning}
On the clients' side, ML model training involves learning the parameters of a Multi-Layer Perceptron (MLP) with three linear layers, producing approximately $12,000$ parameters per client. The learning rates for the IID and non-IID cases are $10^{-5}$ and $10^{-4}$, respectively. The number of epochs is $10$ by default. ReLU activation is applied, and the model is trained using cross-entropy loss with the Adam optimizer.

\subsection{Hardware Setup}

\begin{figure}[htp]
    \centering
    \includegraphics[width=9.5cm]{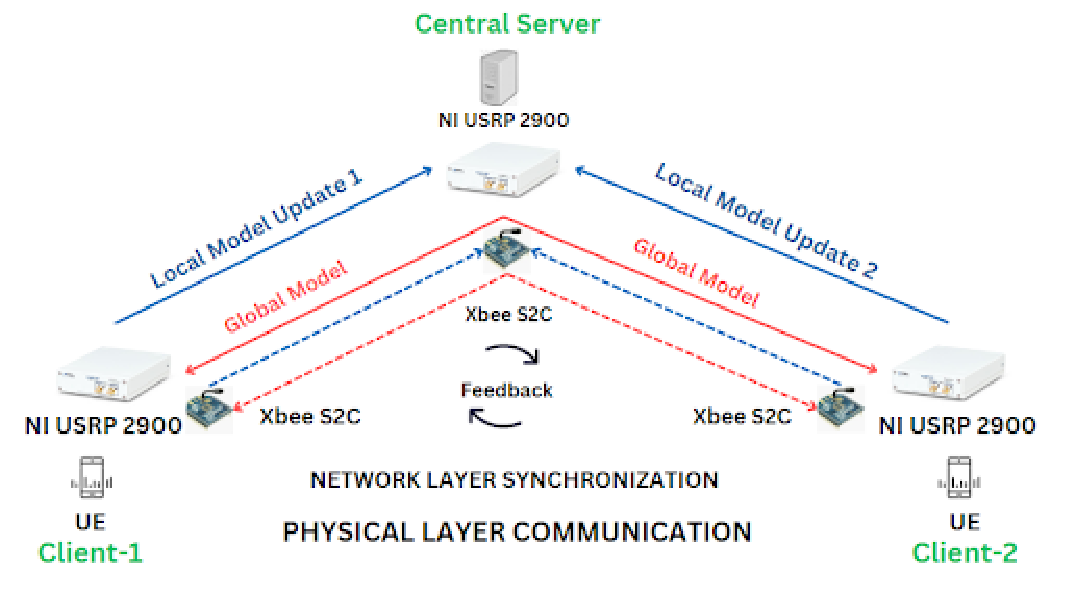}
    \caption{Synergizing Zigbee and USRP for FL Synchronization.}
    \label{fig:ZigbeeUSRP}
\end{figure}

\begin{figure}[htp]
  \begin{minipage}{0.24\textwidth}
      \includegraphics[width=\textwidth, height=1.0\textwidth]{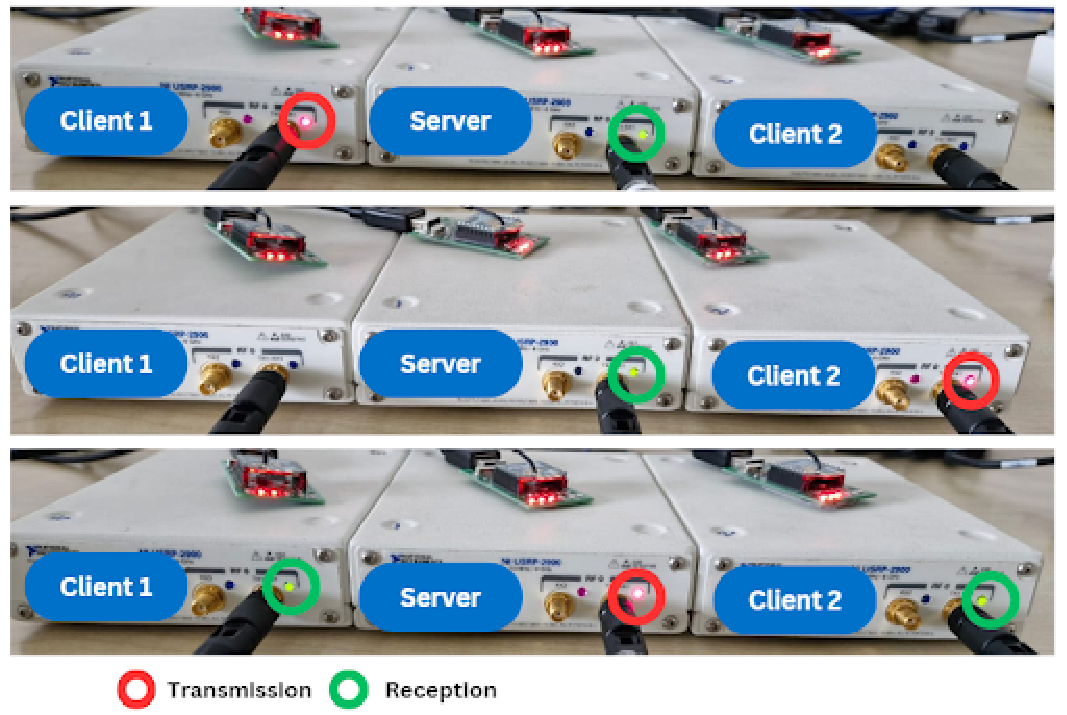}
  \end{minipage}
  \begin{minipage}{0.24\textwidth}
      \includegraphics[width=\textwidth, height=0.9\textwidth]{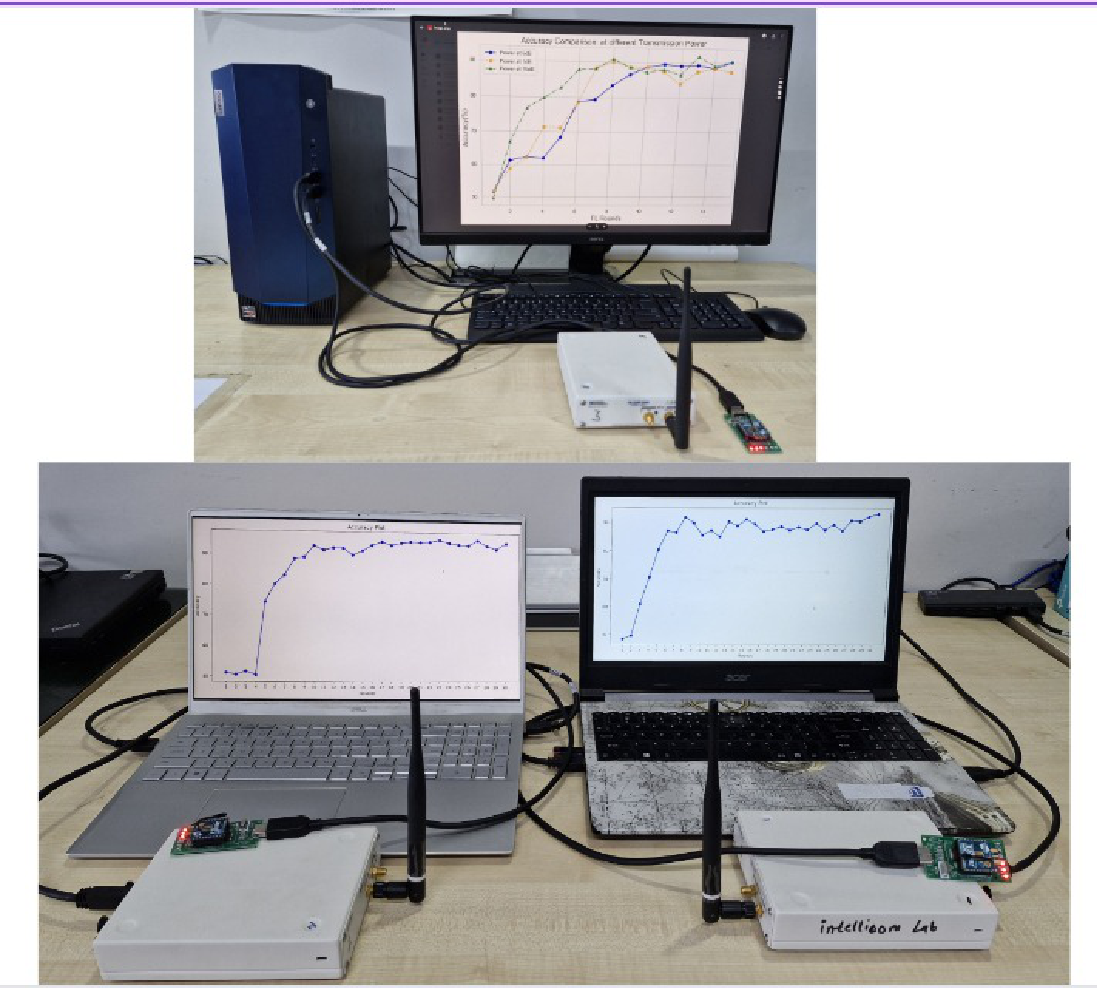}
  \end{minipage}
  \caption{FL prototype using USRPs and Xbee devices.}
  \label{fig:Prototype}
  \vspace{-6mm}
\end{figure}

Our hardware setup is partitioned into three core components: Clients 1 and 2 and a central server as depicted in Fig.~\ref{fig:ZigbeeUSRP}. As shown in Fig.~\ref{fig:Prototype} each of these segments is equipped with essential hardware components, including NI USRP 2900 and Xbee S2C devices. Here, the NI USRP 2900, mounted with an omnidirectional VERT 2450 antenna, is used for physical layer communication, and Xbee S2C devices are used for network-level synchronization. The USRPs leverage differential modulation to enhance system robustness and eliminate the need for channel estimation. The other key signal specifications encompass various parameters crucial for the communication system. These include an operating frequency of $2.5$ GHz, pulse shaping using Root Raised Cosine, a reception gain of $20$ dB, an IQ rate of $50$k, a symbol rate of $6.25$k with $8$ samples per symbol, and transmission power of $20$dB by default. 

%These specifications collectively define the characteristics and performance metrics of the signal within our system.
% Key signal specifications include:
% \begin{itemize}
%     \item Operating Frequency: 2.5 GHz
%     \item Antenna: Omnidirectional VERT 2450
%     \item Pulse Shaping: Root Raised Cosine
%     \item Reception Gain: 20 dB
%     \item IQ Rate: 50k
%     \item Samples per Symbol: 8
% \end{itemize}

\subsection{Synchronization Process}

\begin{figure}[h]
    \centering
    \includegraphics[width=0.8\linewidth] {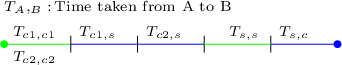}
    \caption{Temporal aspects of one communication round.}
    \label{fig:Time}
    \vspace{-5mm}
\end{figure}

The synchronization process is a challenging aspect of implementing a wireless FL system since the transmission of parameter updates needs to be scheduled meticulously. We realize synchronization by employing Time Division Multiple Access (TDMA) to optimize the scheduling of transmissions. It commences with Client 1, transmitting its locally trained model parameters to the server (in $T_{c1,s}$ seconds). Upon receiving this data, the server promptly acknowledges the reception through the established Zigbee network. It is important to note that the server intentionally waits for a response from Client 1 before the server signals Client 2 to initiate the transmission of its locally trained model parameters. Following the reception of data from Client 2 (in $T_{c2,s}$ seconds), the server, once again, acknowledges the successful reception through the Zigbee network. Subsequently, with data from both clients in hand and having received confirmation responses from both clients via the Zigbee network, the server calculates the necessary aggregate of local model parameters to update the global model. Subsequently, the server broadcasts a signal to both clients, instructing them to activate their receptors. The synchronization process concludes one communication round when both clients independently confirm to the server through the Zigbee network that they have received the updated information, at which point the server halts its broadcasting. The order of activation of the SDRs has been depicted in Fig.~\ref{fig:Prototype} (left), and the overall time taken for such a TDMA-based scheme is shown in Fig.~\ref{fig:Time}. Note that $T_{c1,c1}$, $T_{c2,c2}$ and $T_{s,s}$ refer to computation time at client1, client 2 and the server, respectively.

\subsection{Client-Server Communication}
In our demonstration, USRP is managed through LabVIEW, the XBee devices are controlled using XCTU, and the ML model parameters are generated via Python. These software components seamlessly work in sync, creating an integrated system for the FL experiment.
After obtaining the model updates from the local ML models, it is converted into binary format. This binary data is subsequently organized into packets marked by a unique identifier, effectively serving as its `jacket number.' These packets consist of $128$ bits and are further enhanced by the inclusion of guard bits, sync bits (PN Sequence order $=31$), as well as blank frames. A $60$-bit cyclic redundancy is added to each packet to bolster recognition at the receiver's end. Once these packets are prepared, the final data packet undergoes differential modulation, transforming it into a complex signal (I+Q). This complex signal then advances through the up-conversion stage, ultimately culminating in the process of final transmission.
After down-converting the sampled signal over an acquisition time of $40$ milliseconds, we eliminate any DC offset and unwanted frequencies by applying the Hamming window. Subsequently, we employ threshold voltage detection to transform the analog signal into a digital bit stream through demodulation. The synchronized detection is facilitated through convolution using the same PN sequence order. Once the synchronization process is complete, we extract the message bits from the data packets. These bits are then mapped to their corresponding packet numbers embedded into the generated bit stream.

%Upon successful retrieval of all the required data packets, the system initiates communication with the connected Zigbee device, requesting the transmission of a confirmation response.

\section{Numerical Results}

\begin{figure}
  \begin{minipage}{0.24\textwidth}
      \includegraphics[width=\textwidth, height=0.8\textwidth]{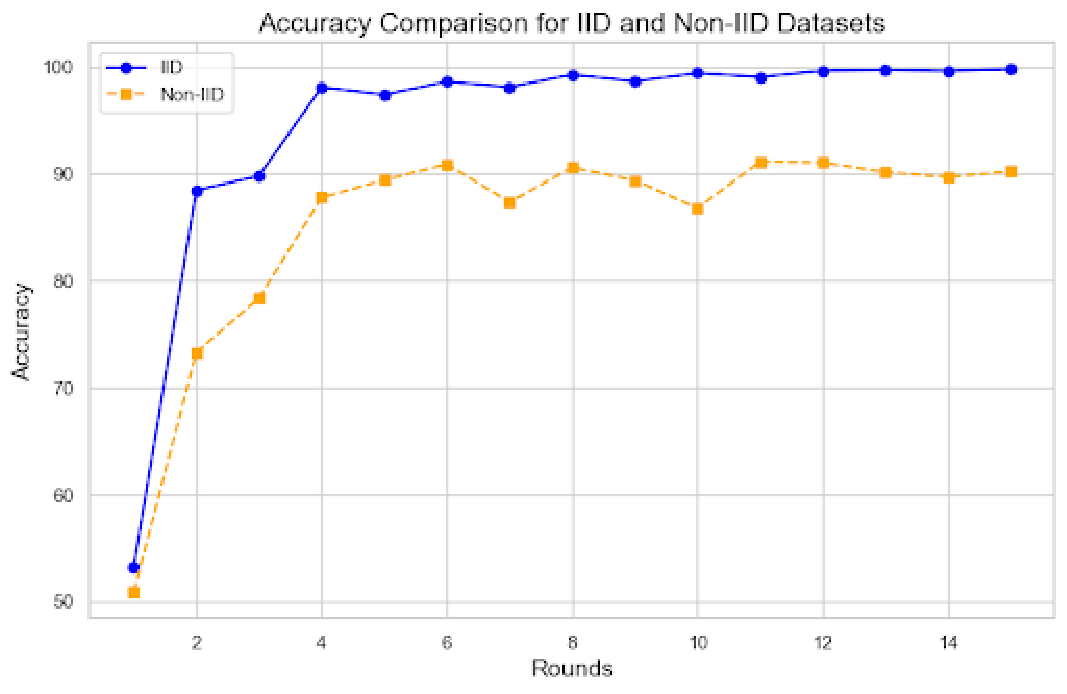}

  \end{minipage}
  \begin{minipage}{0.24\textwidth}
      \includegraphics[width=\textwidth, height=0.8\textwidth]{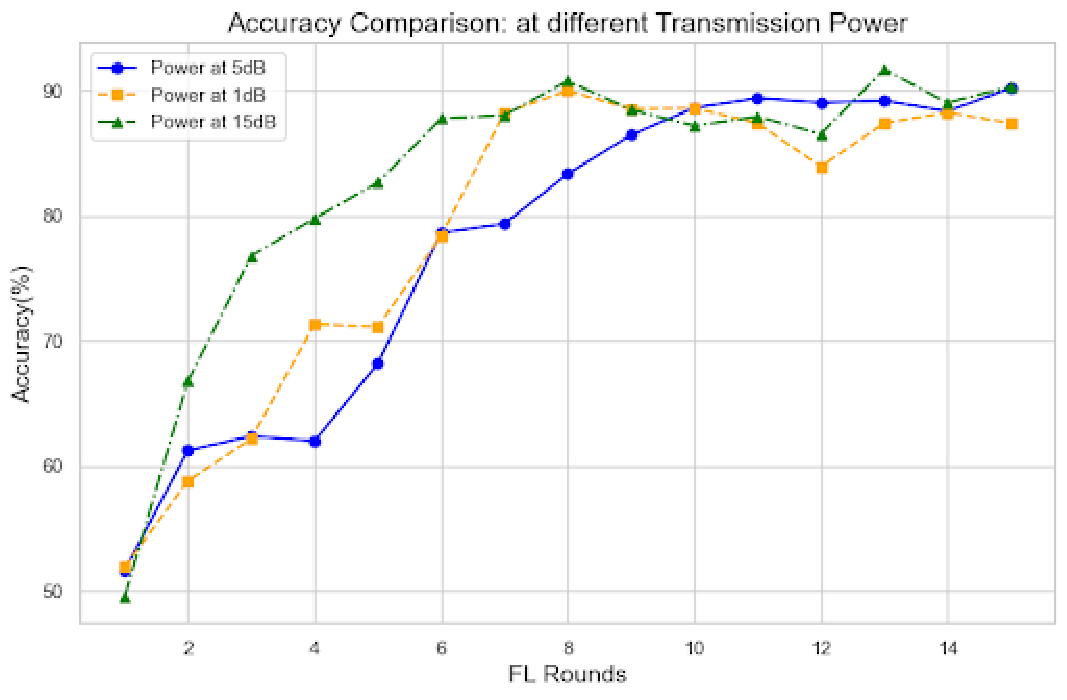}

  \end{minipage}
  \caption{(Left)Accuracy performance in the IID and non-IID setting \& (right) for different transmission powers.}
  \label{fig:accplots}
  \vspace{-5mm}
\end{figure}

In this section, we present the outcomes of our demonstration. In Fig.~\ref{fig:accplots} (left), we compare IID and non-IID scenarios comprehensively. While we see that in the IID scenarios, an accuracy of $99\%$ is obtained, it drops to $90\%$ under severe heterogeneity, as expected. From Fig.~\ref{fig:accplots} (right), we see that transmission power impacts the stability, accuracy, and rate of convergence in the non-IID setting. However, we also observe improved convergence with higher transmission power.

\begin{figure}[htp]
    \centering
    \includegraphics[width=0.98\linewidth, height=0.15\textheight]{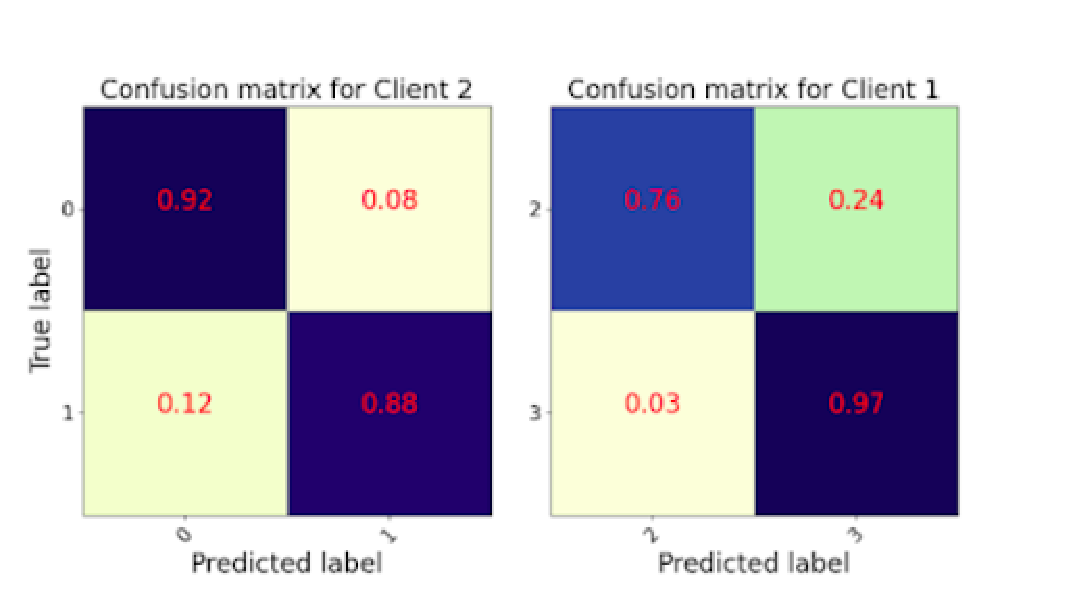}
    \caption{Confusion Matrix for Cross-accuracy.}
    \label{fig:ConfusionMatrix}
\end{figure}

In Table.~\ref{tab:results}, we report the best accuracy as a function of the epochs, which measures the amount of local learning and the transmission power. We observe that the convergence is fast in the case of IID data, and higher local learning helps. Albeit reduced accuracy gains in the non-IID context, more local learning helps here as well. However, the time taken to achieve convergence is much larger as it takes several more communication rounds as compared to the IID setting.
    
In order to ensure that we have a generalized global model, we test the cross-accuracy, i.e., since Client 1 has classes $0$ and $1$, we test if Client 1 can predict well on classes $2$ and $3$, and analogously, we test Client 2 on classes $0$ and $1$. In Fig.~\ref{fig:ConfusionMatrix}, the confusion matrix shows that the cross-accuracy reported by Client 1 is $87\%$ and Client 2 is $90\%$, which ascertains that our global ML model is well-generalised.
    
\begin{table}[h]
\centering
\caption{Best accuracy values for varying epochs and transmission power along with the time taken.}
\label{tab:results}
\begin{tabular}{|c|c|c|c|c|}
\hline
\multirow{2}{*}{\textbf{IID}} & \multicolumn{2}{c|}{\textbf{\# Epochs}} & \multicolumn{2}{c|}{\textbf{Power(dB)}} \\
\cline{2-5}
& \textbf{5} & \textbf{10} & \textbf{10} & \textbf{15} \\
\hline
Accuracy(\%)  & 99  & 99 & 99  & 99\\
\hline
\# Tx. Rounds & 5 & 3 & 3 & 3\\
\hline
Time taken(s) & 215.43 & 180.89 & 180.89 & 178.71 \\
\hline
\multirow{2}{*}{\textbf{Non-IID}} & \multicolumn{2}{c|}{\textbf{\# Epochs}} & \multicolumn{2}{c|}{\textbf{Power(dB)}} \\
\cline{2-5}
& \textbf{5} & \textbf{10} & \textbf{10} & \textbf{15} \\
\hline
Accuracy(\%)  & 91  & 90  & 90  & 92\\
\hline
\# Tx. Rounds & 11 & 8 & 8 & 8\\
\hline
Time taken(s) & 539.55 & 507.36 & 507.36 & 494.63 \\   
\hline
\end{tabular}
\end{table}

\vspace{-2mm}
\section{Conclusions}

\noindent In this work, we present a hardware prototype of FL system that utilizes a wireless backbone. The results of our research exemplify the power of FL in establishing a collaborative and privacy-preserving learning environment over wireless channels where multiple clients collectively contribute to a unified, general model. Establishing a network of SDRs and USRPs to enable private ML, while achieving performance similar to centralized architectures is the key takeaway of this prototype. Our approach offers a basic blueprint for implementing FL, which can be extended to delay-specific wireless networks, network with stragglers, etc.


\vspace{-2mm}

\begin{thebibliography}{00}
\bibitem{b1} B. McMahan, E. Moore, D. Ramage, S. Hampson, and B. Aguera y Arcas. ``Communication-efficient learning of deep networks from decentralized data." In AISTATS 2017, pages 1273–12
\bibitem{b2}T. Sery, N. Shlezinger, K. Cohen and Y. C. Eldar, ``COTAF: Convergent Over-the-Air Federated Learning," GLOBECOM 2020, pp. 1-6.
\bibitem{b3} Amiri, M., and D. Gündüz. "Federated learning over wireless fading channels." IEEE Trans. on Wireless Communications 19, no. 5 (2020).
\bibitem{b4} Hussain, A., N. Gundapu, S. Drugkar, S. Kiran, J. Harshan, and R. Prasad. "Seeing is Believing: A Federated Learning Based Prototype to Detect Wireless Injection Attacks." arXiv preprint:2311.06564 (2023).
\end{thebibliography}
\end{document}